\documentclass[ reprint,onecolumn,12pt, nofootinbib]{revtex4-1}
\usepackage{graphicx}\usepackage{color}
\usepackage{amsmath}

\def\bea{\begin{eqnarray}}
\def\eea{\end{eqnarray}}
\begin{document}

\title{Diffusion of dark matter in a hot and dense nuclear environment}

\author{Marina Cerme\~no $^1$~\footnote{marinacgavilan@usal.es}, M. \'Angeles P\'erez-Garc\'ia$^1$~\footnote{mperezga@usal.es} and Joseph Silk$^{2, 3, 4}$~\footnote{silk@iap.fr}}

\affiliation{$^1$ Department of Fundamental Physics, University of Salamanca, Plaza de la Merced s/n 37008 Spain\\ $^2$Institut d'Astrophysique,  UMR 7095 CNRS, Universit\'e Pierre et Marie Curie, 98bis Blvd Arago, 75014 Paris, France\\ $^3$Department of Physics and Astronomy, The Johns Hopkins University, Homewood Campus, Baltimore MD 21218, USA\\
$^4$Beecroft Institute of Particle Astrophysics and Cosmology, Department of Physics, University of Oxford, Oxford OX1 3RH, UK,\\
}

\date{\today}

\begin{abstract}

We calculate the mean free path in a hot and dense nuclear environment for a fermionic dark matter particle candidate in the $\sim$GeV mass range interacting with nucleons via scalar and vector effective couplings. We focus on the effects of density and temperature in the nuclear medium in order to evaluate the importance of the final state blocking in the scattering process. We discuss qualitatively possible implications for opacities in stellar nuclear scenarios, where dark matter may be gravitationally accreted.
\end{abstract}

\maketitle

\section{Introduction}

There are multiple indications pointing toward a model containing cold dark matter (DM)  as the best explanation for the universe we see at different scales. In particular, the importance of the dark sector component of matter on stellar scales has been less extensively studied and has mainly focused on the sun, planets, white dwarfs \cite{wddm, sun, peter} and compact stars \cite{nsdm}. This is due to the expected moderate  capability of gravitational accretion by these individual celestial bodies from an existing DM galactic distribution. 
Although the current situation of DM searches has greatly benefited from a world-wide experimental effort, at the present  time there still remains a relatively vast DM phase space to explore. Regarding possible values for DM particle masses within the weak interaction sector ({\it wimp} scenario), most popular candidates range from  the sub-GeV region up to $\sim$ 100 TeV. As for the interaction cross-sections with nuclear matter (i.e. nucleons, $N$), there are at least five orders of magnitude in the $\sim$ GeV mass range remaining to be fully tested, namely $\sigma_{\chi N}\sim 10^{-43}$-$10^{-48}$ cm$^2$ as quoted by direct detection searches \cite{review}. 

Apart from this, concerning the nature of the DM particle, in the case of a Majorana candidate, the expected  indirect signal involving gamma-rays or neutrino final products is still under debate \cite{indirec}. In this same direction we may cite other astrophysical effects such as the modification of the emissivity of Standard Model neutrinos from solar reaction chains that have been recently suggested \cite{lopes-silk}.  In the case of asymmetric candidates, accretion of DM mass beyond a critical value, i.e. the Chandrasekhar mass, could induce a dramatic fate for the star where DM accumulates over time \cite{kouv,zurek,bramante} and eventually collapses to a black hole. Another catastrophic event could be triggered following compact object formation via DM seeding. In case of a Majorana candidate, it could induce spark formation energetic enough to  nucleate stable bubbles of deconfined quark matter leading to a softening of the nucleon equation of state. This would drive  a neutron star to quark star conversion \cite{perez-silk1, perez-silk2, perez-silk3}. In addition, unstable DM can also be constrained by structural stability of  accreting objects \cite{perez4}.

However, aside from a pure particle physics description, from the thermodynamical point of view, average magnitudes incorporating the effect of a novel dark sector could be, in principle, determined by evaluating the interplay of both types of matter in a common environment. Typically, the possible dark self-interaction effects are expected to be small as long as the numbers of DM particles remain tiny at all times, with respect to the baryons, and their  relative fraction $Y_\chi=N_\chi / N_B \ll 1$. This could be important, however, for a precise determination of the critical dark matter mass capable of being sustained in a star \cite{kouv}.

As mentioned before, for a given candidate, $\sigma_{\chi N}$ mainly determines the relative fraction of DM to be captured by a compact-sized (spherical) object of mass $M$ and radius $R$. Once inside, it is believed to diffuse toward the denser central stellar regions according to the exponential law $\sim e^{-m_\chi \phi(r)/k_BT(r)}$, being $T(r)$ a local temperature, $\phi(r)$ the gravitational potential and $r$ the radial coordinate \cite{ritz}. 

As an order-of-magnitude estimate, the mean free path of a DM particle, $\lambda_\chi$, is quoted as $\lambda_\chi\simeq 1/\sigma_{\chi N} n$ where  $n$ is the ordinary nucleon number density. This is usually considered as being  sufficient to obtain knowledge about the most efficient opaque environments. For example, a dense nuclear medium such as the central core in a neutron star (with a content $\gtrsim 90\%$ neutrons), exhibits densities well in excess nuclear saturation density $n_{0}\simeq 0.17$ $\rm fm^{-3}$. It is important to note, however, that  in-medium effects are mostly absent from the previous rough estimate. Let us briefly comment on some of the missing corrections. To begin with, Fermi-blocking due to partial restriction of the outgoing nucleon phase space can play a role diminishing the ${\chi N}$ cross section. Finite temperature effects will additionally allow the population of higher energy states in the nucleon sector with respect to the vanishing temperature case to provide the opposite effect. Let us remind  ourselves here that temperatures in the range $T\lesssim 50$ MeV are usually achieved in the very early stages of proto-neutron star evolution \cite{page04}. Later, after a primary neutrino cooling era, temperatures fall to the $\sim$ keV range. This will effectively set at large times a $T\approx0$ configuration, as thermal energies are indeed much smaller than nucleon Fermi energies  $k_BT<<E_{FN}$ in the dense medium.

Motivated for the need to compare bounds from the colliders to direct detection, one can describe interactions between DM and fermions with effective operators in the context of effective field theories (EFT).
In direct detection searches, for example, a non-relativistic incoming $\chi$ particle with low Maxwellian velocity, $v/c\ll1$ is considered. However, collider searches can constrain the high energy part of the interaction as particles are increasingly more relativistic $v/c\sim 1$ allowing, in addition, higher values of $\sqrt{s}$ and momentum transfer $\sim$ TeV. Typically, all the quoted interactions have been largely explored in very low density or vacuum conditions. In particular, scalar, vector and pseudoscalar couplings can play a crucial role as seen in \cite{cheng, fermiondm}. To test increasing relativistic velocity ranges, natural sources of acceleration can be provided by gravitational boosting near compact stars. In this way (neutral) DM particles can acquire large velocities $v\sim c$ and scatter very dense macroscopic regions of size nearly the radius of the star $R\sim 10-12$ km. This extent  has so far only been marginally explored \cite{mirco,p2}. In this work we will focus on the impact of the relativistic contribution of scalar and vector  $\chi N$ couplings to the spin-independent (SI) diffusion of DM inside a dense and hot nuclear medium.

The structure of this contribution is as follows. In section II, we present the effective field theory Lagrangian model using dark matter-nucleon contact interaction via  scalar and vector couplings in a relativistic framework. Later, we compute the doubly differential and integrated $\chi N$ cross sections at finite nucleon chemical potential and temperature. We especially focus on the resulting diffusive behavior of weakly interacting DM particles. In Section III we discuss the obtained dependencies by presenting the figures for selected cases. Finally, in Section IV we give our conclusions.

\section{Dark Matter model and cross sections}

We consider a fermionic dark matter particle of Dirac type, $\chi$, with scalar and vector couplings to the nucleon field $N$ (protons and neutrons). We can write the interaction Lagrangian under the form
\begin{equation}
\mathcal{L_I}=\sum_{N=n,p}g_{Ns}\chi\overline{\chi} N \overline{N}+g_{Nv}\chi\gamma^{\mu}\overline{\chi} N\gamma_{\mu} \overline{N},
\label{lagr}
\end{equation}
where $\gamma_{\mu}$ are the Dirac matrices and $g_{Ns}, g_{Nv}$ are the scalar and vector coupling constants, respectively. 
This treatment is already used in direct detection at low energies with non-relativistic effective field theory operators as shown in \cite{fitz}. Typically, elastic scattering (rather than inelastic) is considered as it is the case relevant for direct detection. Generically, operators containing two  fermionic dark matter fields can be categorized as shown in \cite{p1,review}. In particular, we will focus on those labeled D1 and D5, both contributing to the SI interaction.  
This interaction is equivalent to considering a Fermi four-fermion interaction model, where the effective couplings of mass dimension $(-2)$ for these operators are obtained by integrating out the propagator of a generic $\phi$ mediator with mass $M_\phi$. Motivated by the need to compare bounds from colliders  to direct detection, we describe interactions of DM with quarks $q=u,d$ and averaging in terms of nucleon fields we can write for the vector case $g_{Nv}/M_\phi^2\sim 1/{\Lambda_v}^2$ and $g_{Ns}/M_\phi^2\sim m_q/{\Lambda_s}^3$ where ${\Lambda_v}$ ( ${\Lambda_s}$) is the suppression mass scale for the vector (scalar) case. As usual, we are assuming the effective couplings are of order $O(1)$ and can be absorbed into ${\Lambda_{s,v}}$ \cite{ci}. Using bounds from CMS and ATLAS \cite{cms,atlas} we set  ${\Lambda_v}\gtrsim 1$ TeV and ${\Lambda_s}\gtrsim 100$ GeV. At this point it is worth to mention that a larger parameter range can be considered by means of a multiplicative factor in  each coupling, $g_{Ns}, g_{Nv}$. We have selected these values as they refer to families of phenomenological models that are currently allowed.


Usually, the incoming DM particle is supposed to be thermalized in the galaxy with the Maxwellian mean velocities ${\bar v}\sim 220$ $\rm km/s$. However, in the scenario we consider, an accreting dense star (typically with the mass and dimensions of a neutron star), general relativistic effects are non-negligible and are capable of providing a sizable gravitational boost to the incoming DM particle \cite{banados, mirco}. Let us consider, in order to be concrete, a canonical neutron star of mass $M_{NS}\simeq 1.5 M_{\odot}$ and radius $R_{NS}\simeq 12$ $\rm km$. Expliciting the ratio used as unity, $\frac{GM_{NS}}{c^2}=1$, the velocity modulus $v$ at the star surface is given by
\begin{equation}
\beta=\frac{v}{c}= \sqrt{\frac{2GM_{NS}}{rc^2}}\approx 0.6 \sqrt{\left( \frac{12\,\rm km}{R_{NS}} \right) \left(\frac{M_{NS}}{1.5 \,M_{\odot}}\right)},
\end{equation}
yielding a minimum Lorentz factor at the surface $\gamma=1/\sqrt{1-\beta^2}\approx 1.26$. If scattering happens well inside the core, the previous value is a lower limit, then $\gamma \gtrsim 1.26$. The associated  wavelength of the incoming DM particle is $\lambda=\frac{2 \pi \,\hbar c}{\sqrt{\gamma^2 -1}\,m_\chi c^2}$. This expression sets, in practice, a measure of the validity of our calculation since matter is tested to sizes around $\lambda\sim$ 1 fm, i.e. in the DM mass range $m_\chi\lesssim 5$ GeV. Although further modeling would be required for the description of the inner hadron structure, the use of nuclear form-factors can somewhat mitigate the short-range correlations arising in our calculation as we will see later in the manuscript.  

In order to calculate the differential cross-section per unit volume for the  DM-nucleon scattering, we use  the interaction terms appearing in Eq. (\ref{lagr}). We denote $p'^{\mu}=(E',\vec{p'})$ and  $p^{\mu}=(E,\vec{p})$ as the four-momentum for the outgoing and incoming nucleon of effective mass $m^*_N$, respectively, and $k'^{\mu}=(\omega',\vec{k'})$ and $k^{\mu}=(\omega,\vec{k})$ the analogous for the DM particle of mass $m_\chi$. Momentum transfer is denoted by $q^{\mu}=p'^{\mu}-p^{\mu}=k^{\mu}-k'^{\mu}$. In this way $q_{0}=E'-E=\omega-\omega'$ and $\vec{q}=\vec{p'}-\vec{p}=\vec{k}-\vec{k'}$. The general expression can be written as \cite{pdg, reddy}
\begin{equation}
d\sigma=\frac{|\mathcal{\overline{M}_N}|^2}{4\sqrt{(pk)^2-{m^*}_N^2 m_{\chi}^2}}d\Phi(p,p',k,k')\mathcal{F_{FB}}
\label{dsigma}
\end{equation}
where the phase space volume element is
\begin{equation}
d\Phi(p,p',k,k')=(2\pi)^4\delta^{(4)}(p+k-p'-k')\frac{d^3\vec{p'}}{(2\pi)^32E'}\frac{d^3\vec{k'}}{(2\pi)^32\omega'},
\end{equation}
$|\mathcal{\overline{M}_N}|^2$ is the square of the scattering amplitude of the process considered in our interaction model that we will discuss in detail later. The four-dimensional delta assures the conservation of momentum and energy in the collision. The factor $\mathcal{F_{FB}}$ accounts for the Fermi blocking term that takes into account the occupation of states and  in our calculation affects only to the the nucleon sector (protons or neutrons) $\mathcal{F_{FB}}=f_N(E)(1-f_N(E'))$ with $f_{i}(E)=\frac{1}{1+e^{({E-\mu^*_i})/{k_{B}T}}}$ $i=$p,n. $\mu^*_i$ is the effective nucleon chemical potential for a particle with isospin of ith-type. From this point and in what follows we will consider $\hbar=c=1$. Let us remark that for the dark sector, we will assume that all outgoing DM particles states are in principle  allowed and $1-f_\chi(\omega')\approx 1$ since the fraction of DM inside the star remains tiny at all times. The validity of this approximation is given by the estimate of DM particles inside the object as $N_\chi(t)\approx N_{0\chi}+C_\chi \delta t$. Using $C_{\chi} \simeq 6 \times 10^{25} \left(\frac{M}{1.5 M_{\odot}}\right) \left(\frac{R}{12\,\rm km}\right) \left(\frac{1\, \rm GeV}{m_{\chi}}\right)\left(\frac{\rho^{ambient}_{\chi}}{0.3 \rm \frac{GeV}{cm^3}}\right) \left( \frac{\sigma_{\chi N}}{\sigma_0}\right)\,\,\rm s^{-1}$ \cite{kouv} and the number of nucleons in the star  $N_N=N_B\simeq 10^{58}$ we obtain $Y_{\chi}=N_\chi/N_N<10^{-20}$ for an old neutron star with a lifetime $\delta t \sim 10^{6}$ yr. We will assume a cross-section $\sigma_{\chi N}>{\sigma_0}$, larger than the geometrical or critical cross-section \cite{sun} ${\sigma_0}\simeq \pi R^2_{NS} m_N/M_{NS} \sim 10^{-45}$ $\rm cm^2$ so the star can effectively scatter and capture DM. Let us mention here that the cross-section ratio  $\sigma_{\chi N}/{\sigma_0}$ could be, in principle, even smaller than unity but in that case the scattering scenario we present would be mostly insensitive to dark matter. Assuming cross-sections compatible with the range of currently allowed experimental constraints, $\sigma_{\chi N}/{\sigma_0}>1$, however.
DM population in the NS (after the supernova explosion) should not be negligible since the massive progenitor $8M_\odot \lesssim M_{progenitor}\lesssim 15 M_\odot $ can be efficient in the DM accretion process  \cite{perez4}. Then $N_{0\chi}\lesssim 10^{39}$ for an environment with ambient DM density  $\rho^{ambient}_{\chi}\simeq 0.3$ $\rm GeV/cm^3$.  Effective values of nucleon mass and chemical potential define the {\it quasi-particle} nature of the nucleon in the medium and differ from the nude values by the presence of average meson fields. In this work we will consider this correction as obtained in the existing literature and refer for further reading to, for example, \cite{wal}.

Since we are interested in calculating the DM particle mean free path we will also consider the differential and integrated cross-section {\it per unit volume} and thus we must integrate over the incoming nucleon phase space \cite{reddy}. Then our expression reads

\begin{equation}
\frac{d\sigma(\omega)}{V}=\frac{1}{(2\pi)^5}\int d^3\vec{p} \int d^3\vec{k'} \delta(E+\omega -E-\omega')\frac{|\mathcal{\overline{M}_N}|^2}{16\omega' E'\sqrt{E^2\omega^2-{m^*}_N^2m_{\chi}^2}}\mathcal{F_{FB}},
\label{dsigma1}
\end{equation}
where we have performed a partial integration over 3-dimensional  momentum space. The flux expression appearing in the denominator in Eq. (\ref{dsigma1}) as well as the scattering amplitude we will discuss later in the manuscript have, in general, momentum dependences. In the cross-section calculation, we have retained only the lowest order terms following \cite{reddy} since $v^2_{\chi}\sim v^2_N\ll1$ given $\frac{|\vec{p_i}|}{E_i}=v_i$, $i=\chi, N$ from reference values $v_{\chi}\sim 0.6$ and nucleon Fermi velocities $v_N\sim v_{FN}=|\vec{p}_{FN}|/E_{FN}\sim 0.4$ at  $n=\frac{|\vec{p}_{FN}|^3}{3 \pi^2}= n_{0}$. In particular for the flux, this leads to the final expression $\sqrt{(pk)^2-m_N^2m_{\chi}^2}=\sqrt{(E\omega-\vec{p}\vec{k})^2-m_N^2m_{\chi}^2}\simeq  \sqrt{E^2\omega^2-m_N^2m_{\chi}^2}$.

Let us further rewrite Eq. (\ref{dsigma1}), using a dispersion angle $\theta$ for the outgoing DM particle. In this way we obtain $d^3\vec{k'}=|\vec{k'}|^2 2\pi d(cos\, \rm \theta)$ $d| \vec{ k'} | = 2 \pi |\vec{q}| \frac{\omega'}{|\vec{k}|} d | \vec{q} |dq_{0}$. This follows from  $\omega'd\omega'=|\vec{k'}|d|\vec{k'}|$ and $d(cos\, \theta)=\frac{|\vec{q}|d|\vec{q}|}{|\vec{k}||\vec{k'}|}$. Finally, we obtain after a trivial partial integration,
\begin{equation}
\frac{d\sigma(\omega)}{V}=\frac{1}{(2\pi)^4} \int d^3\vec{p} \int d|\vec{q}| \int dq_{0} \delta(q_{0}+E-E')\frac{|\vec{q}|}{|\vec{k}|}\frac{|\mathcal{\overline{M}_N}|^2}{16E'\sqrt{E^2\omega^2-{m^*}_N^2m_{\chi}^2}}\mathcal{F_{FB}}.
\label{eq1}
\end{equation}

In this calculation, we will restrict ourselves to temperatures and densities typical for the  thermodynamical evolution of the stellar core region, that is $T \lesssim 50$ MeV and $n\simeq (1-3)n_{0}$. Due to the fact that squared Fermi nucleon velocities are $v^2_{FN} \ll 1$  we will perform an expansion of the single particle energies for the incoming and outgoing nucleon states
\begin{equation}
E=m^*_{N}+\frac{|\vec{p}|^2}{2m^*_{N}}, \; \;  E'=m^*_{N}+\frac{|\vec{q}+\vec{p}|^2}{2m^*_{N}}.
\end{equation}
In order to perform the integral in Eq.(\ref{eq1}) we express the energy delta function as 
\begin{equation}
\delta(q_{0}+E-E')=\frac{m^*_{N}}{|\vec{p}||\vec{q}|} \delta(cos\, \theta -cos\, \theta_{0}) \Theta(|\vec{p}|^2-|\vec{p}_-|^2),
\label{eqdelta}
\end{equation}
where
\begin{equation}
cos\; \theta_{0}=\frac{m^*_{N}}{|\vec{p}||\vec{q}|}\left( q_{0}-\frac{|\vec{q}|^2}{2m^*_N}\right) ,
\label{cos}
\end{equation}
and 
\begin{equation}
|\vec{p}_-|^2=\frac{{m^*_N}^2}{|\vec{q}|^2}\left( q_{0}-\frac{|\vec{q}|^2}{2m^*_N}\right)^2.
\end{equation}
Let us now discuss the range of the integration variables. For the energy transfer range $-\infty<q_0<\omega-m_{\chi}$, since $m_{\chi}<\omega'<\infty$, and $|\vec{k}|-|\vec{k'}|<|\vec{q}|<|\vec{k}|+|\vec{k'}|$ from the constraint of a real-valued angle. Instead, at  $T=0$, the energy transfer can not be negative, so that $q_0>0$ and  $0<q_0<\omega-m_{\chi}$ and  for the incoming nucleon it follows $|\vec{p}_-|<|\vec{p}|<\infty$. Note that for $T=0$ in the limit of vanishing kinetic energy for the incoming DM particle, the $q_0$ range reduces to zero as the outgoing states are all occupied, therefore providing a null cross-section, while this is not true in the finite $T$ case as more channels are available.

Finally, if we are interested in the doubly differential cross-section, this can be obtained as 
\begin{equation}
\frac{1}{V}\frac{d\sigma}{d\Omega dq_0}=\frac{1}{(2\pi)^4} \int_{|\vec{p}_-|}^{\infty}\frac{d|\vec{p}||\vec{p}|}{4E'}\frac{m^*_N|\vec{k'}|}{|\vec{q}|}\delta(cos\; \theta-cos\; \theta_0)\Theta(|\vec{p}|^2-|\vec{p}_-|^2)\mathcal{M} ,
\label{difsigma}
\end{equation}
with
\begin{equation}
\mathcal{M}=\frac{|\mathcal{\overline{M}_N}|^2 f_N(E)(1-f_N(E'))}{4\sqrt{E^2\omega^2-{m^*}_N^2m_{\chi}^2}}.
\end{equation}

The sum of scalar($s$) and vector($v$) contributions from the Lagrangian in Eq. (\ref{lagr}) gives a scattering amplitude $\mathcal{M}_N=\mathcal{M}_s+\mathcal{M}_v$ and therefore
\begin{equation}
|\mathcal{\overline{M}}_N|^2=\frac{1}{4}\sum_{spins} \mathcal{M}_N \mathcal{M}_N^*= |\mathcal{\overline{M}}_s|^2+ |\mathcal{\overline{M}}_v|^2+\frac{1}{2}\sum_{spins}\mathcal{M}_s^*\mathcal{M}_v,
\label{totalM}
\end{equation}
where
\begin{equation}
|\mathcal{\overline{M}}_s|^2 = \nonumber 4g_{Ns}^2(p'p+{m^*_N}^2)(k'k+m_\chi^2),
\end{equation}
\begin{equation}
|\mathcal{\overline{M}}_v|^2 = \nonumber 8 g_{Nv}^2[2{m^*_N}^2m_{\chi}^2-{m^*_N}^2k'k-m_{\chi}^2p'p+(p'k')(pk)+(p'k)(pk')],
\end{equation}
and
\begin{equation}
\frac{1}{2}\sum_{spins}\mathcal{M}_s^*\mathcal{M}_v = \nonumber 8g_{Ns}g_{Nv}m^*_Nm_{\chi}(pk+pk'+p'k+p'k').
\end{equation}
As a further correction at short ranges, we can model the structure of the nucleon with a form factor $F(|{\vec q}|)$. We will consider a monopolar form with a cut-off parameter $\Lambda=1.5$ GeV. Then  we will replace $g_{Ns}\rightarrow g_{Ns}F(|{\vec q}|^2)$ and $g_{Nv}\rightarrow g_{Nv}F(|{\vec q}|^2)$ with  $F(|{\vec q}|^2)=\frac{\Lambda^2}{\Lambda^2+q^2}$ so that $F(0)=1$. \\

Retaining the lowest order in particle velocities in the averaged squared matrix element we obtain
%
%
\begin{eqnarray}
|\mathcal{\overline{M}}_N|^2 &\simeq & \nonumber 4g_{Ns}^2(E'E+{{m^*}_N}^2)(\omega'\omega+m_\chi^2)+8 g_{Nv}^2(2{m^*_N}^2m_{\chi}^2-{m^*}_N^2 \omega \omega'-m_{\chi}^2 E'E+2E'\omega'E \omega)\\   &+& \nonumber  8g_{Ns}g_{Nv}m^*_Nm_{\chi}(E\omega+E\omega'+E'\omega+E'\omega').
\end{eqnarray}
Let us note that if finite temperature is considered, detailed balance factors must be added  to the medium response to weak probes \cite{paper6, horo} under the form
\begin{equation}
S(q_0,T)=\frac{1}{1-e^{-\frac{|q_0|}{k_{B}T}}} .
\label{fbalance}
\end{equation}
This factor provides the relation between the dynamical nuclear structure factor for positive and negative energy transfers $q_0$ as the thermodynamic environment can donate energy to the outgoing particle. 

As we are interested in  obtaining the total integrated cross-section per unit volume $\frac{\sigma(\omega)}{V}$ and the inverse of it, i.e. the  mean free path, $\lambda_\chi=\left(\frac{\sigma(\omega)}{V}\right)^{-1}$,  we must integrate over all possible outgoing energy transfer values and solid angle. In this way we obtain 
\begin{equation}
\lambda^{-1}_\chi=\frac{\sigma(\omega)}{V}=\frac{m^*_N}{4(2\pi)^3}\int_{0}^{\omega-m_{\chi}} dq_0  \int_{|\vec{k}|-|\vec{k'}|}^{|\vec{k}|+|\vec{k'}|} d|\vec{q}|\int_{|\vec{p}_-|}^{\infty}d|\vec{p}|\frac{|\mathcal{\overline{M}_N}|^2|\vec{p}|f_N(E)(1-f_N(E'))S(q_0,T)}{4E'|\vec{k}|\sqrt{E^2\omega^2-{m^*_N}^2m_{\chi}^2}}.
\label{sigma}
\end{equation}
\section {Results}
In this section we present the results. We start by discussing the low T regime. We set  $g_{Ns} \sim 10^{-15} \; \rm MeV^{-2}$ and $g_{Nv}\sim 10^{-13} \; \rm MeV^{-2}$ and a cut-off parameter $\Lambda=1.5$ GeV. In order to include the effect of  the medium we replace vacuum nucleon mass and chemical potential values by the effective ones at each baryonic density and T \cite{wal}. The $ith$-type isospin is obtained according to $n_i= \frac{2}{(2 \pi)^3}\int_0^{\infty} \frac{4 \pi p^2 dp} {1+e^{\left( \sqrt{p^2+{m^*_N}^2}-\mu^*_i \right)/k_B T}}$.

\begin{figure}[h]
\begin{center}
\includegraphics [angle=0,scale=1.5] {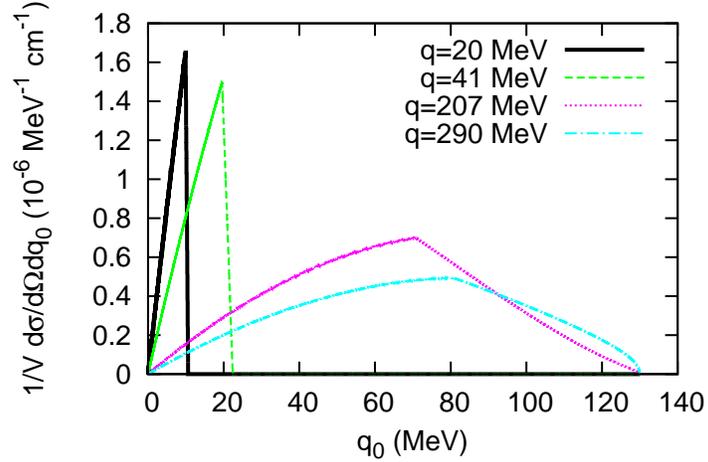}
\caption{Differential cross section per unit volume as a function of the energy transfer $q_0$ for values of $|\vec q|=20, 41, 207$ and $290$ MeV. The DM particle mass is $m_\chi=0.5$ GeV and $T=0$ at $n=n_{0}$.}
\label{fig1}
\end{center}
\end{figure}

 In  Figure~\ref{fig1} we show the differential cross-section per unit volume as a function of the energy transfer $q_0$ for different values of $|\vec q|=20, 41, 207$ and $290$ MeV with dash-dotted, dashed, dotted and solid lines, respectively, for a pure neutron system with $n=n_{0}$. We use $m_\chi=0.5$ GeV setting $T=0$. The limiting upper value of the energy transfer is $\omega-m_\chi \approx 130$ MeV. The triangular shape is due to the Heaviside Fermi distribution at $T=0$.  Beyond $q_0$ values limited by real-valued angles in Eq.(\ref{cos}) the scattered states are not allowed since it is kinematically impossible to scatter a nucleon due to lack of empty states.

\begin{figure}[ht]
\begin{center}
\includegraphics [angle=0,scale=1.5] {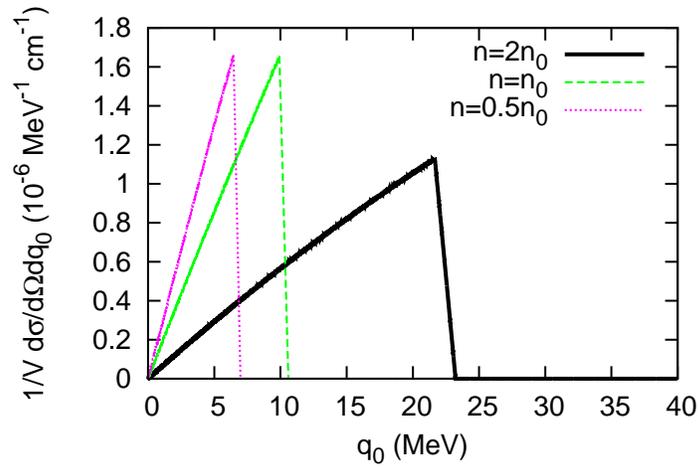}
\caption{Differential cross-section per unit volume as a function of the energy transfer $q_0$ for nucleon densities $n=(0.5,1, 2)n_{0}$. We set  $|\vec q|=20$ MeV and $m_\chi=0.5$ GeV at $T=0$.}
\label{fig2}
\end{center}
\end{figure}

In  Figure~\ref{fig2} we show the differential cross-section per unit volume as a function of the energy transfer $q_0$ for different values of the nucleon number density $n=(0.5,1,2) n_{0}$ with dotted, dashed and solid lines, respectively for  $|\vec q|=20$ MeV. We use $m_\chi=0.5$ GeV setting $T=0$ and effective nucleon masses $m^*_N/m_N\simeq 0.85, 0.7$ and $0.4$ for the increasing density set. A combined effect of the density dependence of nucleon masses and the nucleon Fermi momentum value provide a rapid increase of the maximum $q_0$ value.  

\begin{figure}[ht]
\begin{center}
\includegraphics [angle=0,scale=1.5] {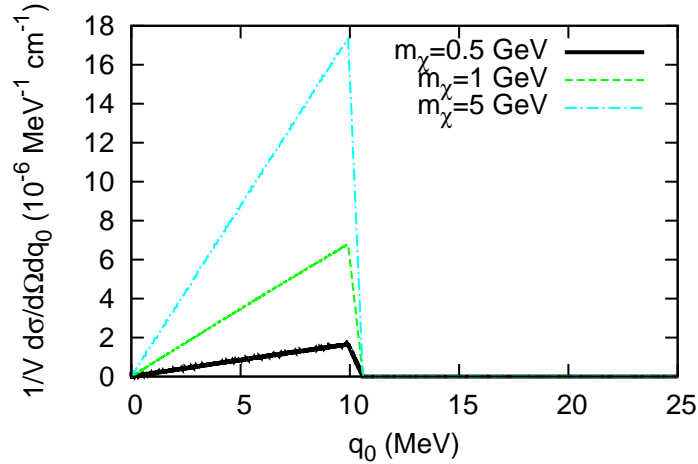}
\caption{Differential cross-section per unit volume as a function of the energy transfer $q_0$ for a nucleon density $n=n_{0}$. We set  $|\vec q|=20$ MeV at T=0. The $m_\chi=5$ GeV case has been decreased a factor 10 in this plot.}
\label{fig3}
\end{center}
\end{figure}

In order to test the variability with the dark probe mass we depict in  Figure~\ref{fig3} the differential cross-section per unit volume as a function of  $q_0$ for mass values $m_\chi=0.5, 1$ and $5$ GeV. We set $T=0$ and $n=n_{0}$ at  $|\vec q|=20$ MeV. The curve with $m_\chi=5$ GeV has been decreased a factor 10 to make the trend more clear.

\begin{figure}[ht]
\begin{center}
\includegraphics [angle=0,scale=1.5] {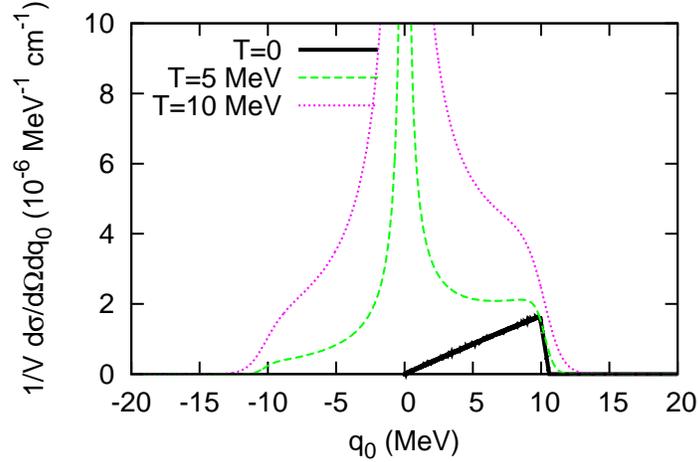}
\caption{Differential cross-section per unit volume as a function of the energy transfer $q_0$ at $T=0, \,5,\, 10$ MeV for a nucleon density $n=n_{0}$. We set  $|\vec q|=20$ MeV and $m_\chi=0.5$ GeV}
\label{fig4}
\end{center}
\end{figure}

Finite temperature effects can be observed in  Figure~\ref{fig4} where the detailed balance factors have been included. We use values of temperature $T=0, \,5$ and $10$ MeV with solid, dashed and dotted lines, respectively for $|\vec q|=20$ MeV. We set a fixed value of the chemical potential $\mu=E_{FN}$ at $n=n_{sat}$. This corresponds to densities $n=0.170, 0.174$ and $0.209$ $\rm fm^{-3}$ setting $m_\chi=0.5$ GeV.
At temperatures $T>0$ the negative energy transfer states get increasingly populated and the sharp nucleon distribution is smoothed. As $q_0 \rightarrow 0$ the inverse detailed balance factor $S^{-1}(q_0, T) \rightarrow 0$. The corresponding divergence will, however, be integrable in order to obtain a finite integrated cross-section.

\begin{figure}[ht]
\begin{center}
\includegraphics [angle=0,scale=1.5] {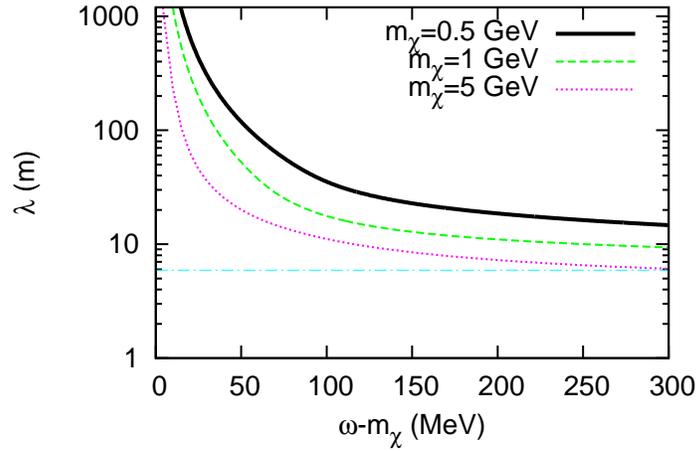}
\caption{Dark matter mean free path as a function of kinetic energy for $m_\chi=0.5, 1$ and $5$ GeV at T=0 and $n=n_{sat}$. Dot-dashed line shows the simplified estimate yields a constant value  $\lambda_\chi\simeq 1/\sigma_{\chi N} n\sim 5.9$ m assuming current experimental sensitivities $\sigma_{\chi N}\sim 10^{-41}$ $\rm cm^2$. See text for details.}
\label{fig6}
\end{center}
\end{figure}
\begin{figure}[ht]
\begin{center}
\includegraphics [angle=0,scale=1.5] {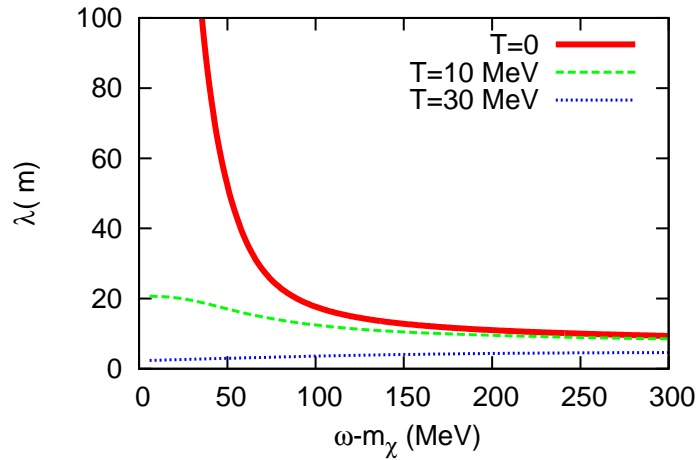}
\caption{DM particle mean free path as a function of kinetic energy for $m_{\chi}=1$ GeV at $n=n_{0}$ for $T=0, 10, 30$ MeV. }
\label{fig8}
\end{center}
\end{figure}

In   Figure~\ref{fig6} we show the mean free path for the $\chi$ particle as a function of kinetic energy $K=\omega-m_{\chi}$ for three mass values  $m_\chi=0.5,1, 5$ GeV with solid, dashed and dotted lines, respectively. We set $n=n_{0}$ and $T=0$. We can see that in this DM mass range, scattering is diffusive to very good approximation as $\lambda/R \ll 1$. We show with dot-dashed line the simplified estimate yields a constant value  $\lambda_\chi\simeq 1/\sigma_{\chi N} n\sim 5.9$ m assuming sensitivities $\sigma_{\chi N}\sim 10^{-41}$ $\rm cm^2$.  For our choice of couplings strengths, fixed $K$ energy, a Standard Model neutrino displays typical mean free path somewhat smaller \cite{reddy,horo,mag} however being an efficient heat carrier inside the star. The larger the mass of the DM particle the more opaque is the medium to it. Note that as $\omega \rightarrow m_{\chi}$ the phase space available for the outgoing particles vanishes as the energy transferred $q_0 \rightarrow 0$. As all the outgoing states are all occupied in the nucleon sea at $T=0,$ this provides a null value of the integrated cross-section value for the DM-nucleon interaction. This behavior is shown in   Figure~\ref{fig8} where we plot the variation of the mean free path with kinetic energies for temperatures $T=0$ (solid line) $T=10$ MeV (dashed line) and $T=30$ MeV (dotted line). We consider $m_{\chi}=1$ GeV and $n=n_{0}$. At $T=0$ and vanishing kinetic energy the mean free path goes arbitrary large as the integrated cross-section also vanishes due to filled population levels. This behavior is smoothed at finite temperature where a non-vanishing mean free path value is recovered. 

\begin{figure}[ht]
\begin{center}
\includegraphics [angle=0,scale=1.5] {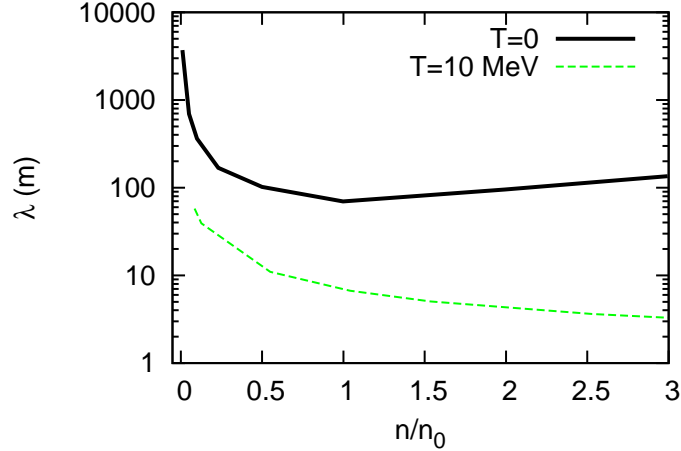}
\caption{DM particle mean free path as a function of density (in units of $n_{0}$) for two values of temperature, T=0 and T=10 MeV. Effective (naked) nucleon mass has been used in the zero (finite) T calculation. }
\label{fig7}
\end{center}
\end{figure}

In   Figure~\ref{fig7} the variation of the DM particle mean free path is shown as a function of density (in units of $n_{0}$) for two values of temperature, T=0 (solid line) and T=10 MeV (dashed line). We use $m_\chi=1$ GeV and effective nucleon masses have been considered for the T=0 case while not for the finite temperature case in order to estimate competitive effects. A steady decrease is obtained in case the naked nucleon mass is considered. Incoming energy has been fixed to $\omega=1.26 m_{\chi}$ for each case. Temperature effects, which are relevant in the early stages of dense star evolution,  tend to increase the opacity of nucleon matter to prevent DM nearly-free streaming. 
 
In order to qualitatively compare our findings with existing current experiments we consider generic sensitivities constrained from direct and collider searches. In this part our aim is to see how our results fit in the global present picture of relativistic scattering of DM coming from a complementary and different scenario.

From our earlier discussion, the scenario we present in our work is meaningful for  $\chi N$ cross-section larger than that of the geometrical cross-section  $\langle\sigma_{\chi N} \rangle  \sim {\sigma_0}\simeq \pi R^2 m_n/M \sim 10^{-45}$ $\rm cm^2$. Typical constrained values in current experiments are larger than this value.  In order to compare strengths we consider a typical reference value of central baryonic density in the star $n=2n_{0}$, and estimate integrated cross-sections {\it per particle}, averaging over the nucleon particle density, as $\langle\sigma_{\chi N}\rangle\simeq \frac{\sigma_{\chi N}/V}{n}$.

Using our set of generic couplings  $g_{Ns},\,g_{Nv}$ we obtain scalar and vector contributions $\langle\sigma_{s,\chi N}\rangle \sim 10^{-47}$,  $\langle\sigma_{v,\chi N} \rangle \sim 10^{-43}$ $\rm cm^2$ for masses in the range $m_{\chi}\sim$GeV. These results must be considered as an averaged value in momentum space and are consistent with existing collider constraints on D1, D5 couplings \cite{atlas,cms} at  $m_{\chi} \sim$GeV range derived in the context of EFT. Note that these quoted constraints on collider and direct searches should be however taken with caution since they present some already well-known problems, i.e. (over-) under-estimates, and inconsistencies with the thermal relic density for  $m_{\chi}$ ranges outside a   $m_{\chi}\sim (170-500)$ GeV range, see a discussion in, for example  \cite{eft, eft2}.
The energy range that we describe in  the relativistically boosted scenario accounts for center-of-mass energies of $\sqrt{s} \lesssim 6$ GeV for  DM candidates with $m_{\chi}\lesssim 5$ GeV scattering target nucleons in a dense nuclear sea. However, as stated in \cite{eft2} it is also the momentum transfer and mediator mass, $M_\phi$, that provide the validity of interpretation of EFT as it must fulfill $q^2 \ll M_\phi^2$.

In the current status of the direct detection DM search in the low mass range region CDMS \cite{cdms} and SuperCDMS SNOLAB \cite{supercdms} provide the best limits up to date with a sensitivity of $\sigma_{\chi N}\sim 10^{-41}$  $\rm cm^2$ for a limiting value $m_{\chi}\sim 5$ GeV. Below this mass, collider searches can provide better sensitivities than direct searches because the momentum transfer becomes small and the nuclear recoil energy falls below experimental thresholds. We expect, nevertheless, that a more refined model of hadron structure or the mediators in the interaction will provide a richer contribution to be determined in the future, as this is far from the present scope of this work. We consider that, despite indirectly, dense astrophysical sites can contribute to probing the low mass region of the DM phase space.

\section{Conclusions}

In summary, we have calculated dark matter scattering cross sections in an environment of dense and hot nucleon matter. We have considered a fermionic DM particle with scalar and vector effective couplings. In this scattering scenario, we have tested  a low mass region $m_\chi \lesssim 5$ GeV. Examples for this setting are the interiors of neutron stars where core densities range typically exceed $n/n_{0}\gtrsim 1$ and temperatures $T\lesssim 50$ MeV. We have included the nuclear medium effects through Fermi-Dirac distributions for the nucleon sector assuming the amount of DM mass in the star remains tiny at all times. To partially correct for the fact that we consider a point-like interaction, we use monopolar form factors for the hadron structure. We find that the differential and integrated cross-sections are greatly affected by the finite density of matter, namely by the effect of a smaller  effective nucleon mass $m^*_N<m_N$. Temperature effects are taken into account with additional detailed balance factors and are found to be  important although to a lesser extent relative to density. The mean free path for a DM particle is found to be larger than the typical values of those found for Standard Model neutrinos with vector-axial couplings.  The simplified estimate for the mean free path , $\lambda_\chi\simeq 1/\sigma_{\chi N} n$, lacks the rich dependence on the phase space of the scattering process. In this paper we show that the diffusive behavior approximation at finite density and temperature in the interior of NS is well grounded and DM can contribute to the energy transport in their interior. While a specific application to proto-neutron stars is deferred to a later paper, here we have discussed the interest of dense neutron stars to expose the importance of the medium effects in the interaction of ordinary and dark matter.

MAPG acknowledges interesting discussions with C. Albertus, R. Lineros and J. Horvath. This research has been partially supported by  University of Salamanca,  FIS2012-30926, FIS2015-65140 and MULTIDARK MINECO projects and at IAP by  the ERC project  267117 (DARK) hosted by Universit\'e Pierre et Marie Curie - Paris 6   and at JHU by NSF grant OIA-1124403. M. Cerme\~no is supported by a fellowship from the  Consolider MULTIDARK project and Universidad de Salamanca.


\end{document}